\documentclass[preprint, superscriptaddress]{revtex4-1}

\usepackage{hyperref}
\usepackage{graphicx}
\usepackage{amstext}
\usepackage{amsmath}
\usepackage{amssymb} 
\usepackage{color}
\usepackage{array}
\usepackage{enumitem}
\usepackage[normalem]{ulem}

\def\be{\begin{equation}}
\def\ee{\end{equation}}
\def\bit{\begin{itemize}}
\def\eit{\end{itemize}}
\def\bal#1\eal{\begin{align*}#1\end{align*}}
\def\ba#1\ea{\begin{align}#1\end{align}}
\def\beq{\begin{eqnarray}}
\def\eeq{\end{eqnarray}}
\def\ra{\rangle}
\def\la{\langle}
\def\nn{\nonumber}
\def\dsp{\displaystyle}
\def\mc{\mathcal}

\def\ep{\epsilon}
\def\h{\hat}

\def\te{\text}
\def\da{\dagger}
\def\bet{\beta}
\def\alp{\alpha}
\def\de{\delta}
\def\ga{\gamma}

\newcommand{\bra}[1]{\langle #1|}
\newcommand{\ket}[1]{| #1\rangle}

\newcommand{\Fig}[1]{Figure~\ref{#1}}
\newcommand{\Eq}[1]{Eq.~\eqref{#1}}
\newcommand{\Eqs}[2]{Eqs.~\eqref{#1} and \eqref{#2}}

\begin{document}
\title{Quantifying fermionic decoherence in many-body systems}
\author{Arnab Kar}
\thanks{Present address: Laboratory for Laser Energetics, University of Rochester, Rochester, New York 14623, USA}
\affiliation{Department of Chemistry, University of Rochester, Rochester, New York 14627, USA}
\author{Ignacio Franco}
\email{ignacio.franco@rochester.edu}
\affiliation{Department of Chemistry, University of Rochester, Rochester, New York 14627, USA}
\affiliation{Department of Physics, University of Rochester, Rochester, New York 14627, USA}

%%%%%%%%%%%%%%%%%%%%%%%%%%%%%%%%%%%%%%%%%%%%%%%%%%%%%%%%%%%%%%%%%%%%%%
\begin{abstract}
Practical measures of electronic decoherence, called distilled purities, that are applicable to many-body systems are introduced. 
While usual measures of electronic decoherence such as the purity employ the full $N$-particle density matrix which is generally unavailable, the distilled purities are based on the $r$-body reduced density matrices ($r$-RDMs) which are more accessible quantities.
The $r$-body distilled purities are derivative quantities of the previously introduced $r$-body reduced purities [I. Franco and H. Appel, J.~Chem.~Phys.~\textbf{139}, 094109 (2013)] that measure the non-idempotency of the $r$-RDMs. Specifically, the distilled purities exploit the structure of the reduced purities to extract coherences between Slater determinants with integer occupations defined by a given single-particle basis that compose an electronic state. In this way, the distilled purities offer a practical platform to quantify coherences in a given basis that can be used to analyze the quantum dynamics of many-electron systems. Exact expressions for the one-body and two-body distilled purities are presented and the utility of the approach is exemplified via an analysis of the dynamics of oligo-acetylene as described by the Su-Schrieffer-Heeger Hamiltonian. Last,  the advantages and limitations of the purity, reduced purity and distilled purity as measures of electronic coherence are discussed.

\begin{flushleft}
A. Kar and I. Franco, J.~Chem.~Phys. \textbf{146}, 214107 (2017)
\end{flushleft}
\end{abstract}
\maketitle

%%%%%%%%%%%%%%%%%%%%%%%%%%%%%%%%%%%%%%%%%%%%%%%%%%%%%%%%%%%%%%%%%%%%%%
\section{Introduction}

Decoherence refers to the change of a state of a system from a pure state $\h{\rho}_e=|\psi\ra\la \psi|$ to a mixed state $\h{\rho}_e=\sum_i w_i |\psi_i\ra\la \psi_i|$ due to interactions with an environment~\cite{Breuer,joos,Schlosshauer,Scholes:2017}.
In isolated electron-nuclear systems, electronic  decoherence arises due to the entanglement of the electrons with the nuclear degrees of freedom caused by electron-nuclear couplings. As such, electronic decoherence is a basic feature of correlated electron-nuclear states~\cite{francoBO} and accompanies most dynamical processes in molecules.
In addition to its interest at a fundamental level, determining mechanisms for electronic decoherence~\cite{Connection_Kar} is central to interpreting
coherence phenomena in matter~\cite{scholes2015, Quantum_Biology,photo3,Collini_Photosynthesis, Prokhorenko_Vision,Scholes:2017}, to the development of methods to follow correlated electron--nuclear dynamics~\cite{Kapral,Prezdho_DISH}, and is the starting point to develop potential protocols to protect coherences for quantum control applications~\cite{Nielsen,quantumcontrol}.

In this paper, we focus on the  problem of how to quantify electronic coherences~\cite{Hwang,Wong,Smyth} in many-body systems by using purity and purity-related measures. 
The purity of an electronic state is defined as 
\be
\label{eq:purity}
P(t)=\text{Tr}[\hat{\rho}_e^2(t)] = \sum_i \lambda_i^2,
\ee
where $\h{\rho}_e(t)$ is the $N$-particle electronic density matrix (obtained by tracing out the nuclear coordinates from the electron-nuclear density matrix $\hat{\rho}$, i.e. $\h{\rho}_e(t)= \text{Tr}_N[\hat{\rho}(t)]$) and $\lambda_i$  its eigenvalues (the Schmidt coefficients squared).   The purity is a well-defined measure of electronic decoherence. It is basis-independent, and easy to interpret ($P=1$ for pure systems where the density matrix is idempotent, $\h{\rho}_e^2 = \h{\rho}_e$, and  $P<1$ for mixed states~\cite{Greiner_expt,purity_expt}).  In spite of these  advantages,  it is often not possible to apply the purity to quantify coherence in many-electron systems because it requires knowing the $N$-particle electronic density matrix $\h{\rho}_e(t)$ which is an experimentally and computationally challenging quantity to obtain except for few-level problems~\cite{densitymatrix_expt}.  

Thus, it is desirable to develop measures of coherence that are based on more accessible quantities such as the $r$-body electronic reduced density matrices ($r$-RDMs), $^{(r)}\hat{\Gamma}$. The $r$-RDMs are defined by tracing out $(N-r)$ electronic degrees of freedom out of $\h{\rho}_e$. In particular, the one and two-body RDMs are useful targets as they can be propagated directly in state-of-the-art simulations of many-body systems~\cite{Lackner},  and they determine  most observable quantities of physical interest. In analogy with the purity [\Eq{eq:purity}], one can define a \emph{reduced purity}~\cite{Franco_Reduced}
\be
\label{eq:rpur}
P_r(t) = \text{Tr}[^{(r)}\hat{\Gamma}^2] =  \sum_i {}^{(r)}\lambda_i^2,
\ee
that measures the non-idempotency of the $r$-RDM, where ${}^{(r)}\lambda_i$ are the eigenvalues of the $r$-RDM. As is the case of purity,  the reduced purities are representation independent and decay with electronic coherence loss in the system \cite{Franco_Reduced}. Nevertheless, contrary to the purity, as a measure of coherence these quantities can be difficult to interpret. This is because the non-idempotency  of the  $r$-RDMs induced by decoherence can also arise due to increased electronic correlation~\cite{Ziesche}.  In fact, due to the reduced information contained in the $r$-RDMs, at the $r$-RDM level it is very challenging  to distinguish contributions of electronic correlation and decoherence to the non-idempotency of $^{(r)}\hat{\Gamma}$.  This makes it difficult to distinguish, for instance, a pure state of an electronically correlated molecule from a mixed state of an electronically uncorrelated molecule. 

As an alternative, here we introduce \emph{distilled purities} [defined by \Eqs{eq:dp1}{eq:dp2}] as a practical measure of fermionic decoherence in many-body systems. The distilled purities are based on the $r$-RDMs and are derivative quantities of the reduced purities. They summarize the coherence content between single Slater determinants that form the electronic  state.  These quantities are easy to calculate and useful in interpreting the quantum dynamics of many body systems in a situation where well-defined measures of decoherence are not accessible. Nevertheless, as discussed below, they have the limitation of being manifestly basis-dependent and of not being simply  related to well-defined measures of coherence such as the purity.  As such, the distilled purities should be seen as useful quantities to interpret dynamics, and not as fundamental quantities. This is akin to atomic population analysis in molecular systems, which are of significant interpretative value but for which there is no apparent correct definition.

This paper is organized as follows: In Sec.~\ref{define}, we define different kinds of coherences that can be associated with a many-body electronic density matrix. As the mathematical formalism of the reduced purities is necessary to introduce the distilled purities, in Sec.~\ref{review} we briefly review the reduced purities and present exact expressions for the one- and two-body reduced purities. Then, in Sec.~\ref{distilledpurity}  we define the distilled purities, derive exact expressions for the one-body and two-body distilled purities and isolate their limiting values.
In Sec.~\ref{results}, the behavior of the reduced and distilled purities is exemplified for a model molecular system with electron-vibrational interactions.
Last, in Sec.~\ref{conclude} we summarize our findings and discuss the merits and limitations of the purity, reduced purity and distilled purity as measures of electronic decoherence.

%%%%%%%%%%%%%%%%%%%%%%%%%%%%%%%%%%%%%%%%%%%%%%%%%%%%%%%%%%%%%%%%%%%%%%
\section{Definitions of coherence}\label{define}

We begin by defining different types of coherences that can be studied in the context of many body systems.  For definitiveness, consider a pure state of an electron-nuclear system in Schmidt form $|\Omega\ra=\sum_i \sqrt{\lambda_i}|\mc{E}_i\ra|R_i\ra$ where $\sqrt{\lambda_i}$ are the Schmidt coefficients, and $|\mc{E}_i\ra$ and $|R_i\ra$ are orthonormal states of the electronic and nuclear subsystems.  In the eigenbasis of the electronic Hamiltonian for a fixed nuclear configuration  \{$|E_j\ra$\},  $|\Omega\ra=\sum_j |E_j\ra|\chi_j\ra$
where $|\mc{E}_i\ra=\sum_j c_{ij} |E_j\ra$ and $|\chi_j\ra=\sum_i \sqrt{\lambda_i}c_{ij}|R_i\ra$. The  $|\chi_j\ra$ can be viewed as the nuclear wave packets associated with the $j$-th electronic state. In this context, the $N$-body electronic density matrix is given by
\be\label{Nbody}
\begin{split}
\hat{\rho}_e(t)&=\te{Tr}_N[|\Omega(t)\ra\la\Omega(t)|]=\sum_i \lambda_i \ket{\mc{E}_i}\bra{\mc{E}_i}\\
&=\dsp{\sum_{j,k} \la\chi_k(t)|\chi_j(t)\ra|E_j\ra \la E_k|},
\end{split}
\ee
where the partial trace is over the nuclear degrees of freedom~\cite{Franco_Reduced}.  In the \{$|E_j\ra$\} eigenbasis, the off-diagonal elements of the electron density matrix are determined by the overlap $|\la\chi_k(t)|\chi_j(t)\ra|$  between the nuclear wavepackets associated with different electronic states.  In turn, in the Schmidt basis $\hat{\rho}_e(t)$ is a diagonal matrix with the squared Schmidt coefficients in the diagonal ($\sum_i \lambda_i = 1$).

\paragraph*{Coherence.} We define the \emph{degree of coherence} of $\hat{\rho}_e(t)$ through the purity
\be\label{Npurity}
\begin{split}
P(t)&=\text{Tr}[\hat{\rho}_e^2(t)]=\sum_i \lambda_i^2 \\
&= \dsp{\sum_{j,k}|\la\chi_k(t)|\chi_j(t)\ra|^2}\leq 1.
\end{split}
\ee
For pure systems, only  one Schmidt coefficient is non-zero and the purity is one. For mixed states with reduced coherence properties, there is more than one  non-zero Schmidt coefficients, and the purity is less than one.  From the perspective of electron-nuclear dynamics, such decoherence occurs due to a decay of the overlaps $|\la\chi_k(t)|\chi_j(t)\ra|$  between the nuclear wavepackets associated with different electronic states. 

\paragraph*{$B$-Coherence.}  The purity definition of coherence is well defined and basis-independent. 
Nevertheless, often  by coherences it is simply meant the non-zero off-diagonal elements of the subsystem's density matrix expressed in a given basis~\cite{Kassal_definition,Mukamel}, typically the position or the energy basis. 
For instance,  the coherences of $\hat{\rho}_e$ in the  \{$|E_j\ra$\} eigenbasis [see \Eq{Nbody}] are given by the nuclear wavepacket overlaps $|\la\chi_k(t)|\chi_j(t)\ra|$.
We will refer to this type of basis-dependent coherences as \emph{$B$-coherences}. 
While a decay in the $B$-coherences can lead to decay in the state purity, the absence of $B$-coherences does not necessarily signal an incoherent state. In this sense, the $B$-coherences are useful in interpreting the dynamics of the system, but they are not necessarily well-defined measures of coherence. 

\paragraph*{S-Coherence.}  
As a particular class of $B$-coherences, in many-fermion systems one is often interested in the $B$-coherences of the electronic density matrix expressed in a basis of single Slater determinants. We will refer to this type of coherence as $S$-coherences. These Slater determinants are defined by a given single-particle orbital basis and refer to the anti-symmetrized products of such orbitals with integer occupation numbers. A complete basis of single Slater determinants is constructed by considering all possible distribution of the electrons among the single-particle states.

These three definitions of coherences will be of importance when discussing the utility of the reduced and the distilled purities.

%%%%%%%%%%%%%%%%%%%%%%%%%%%%%%%%%%%%%%%%%%%%%%%%%%%%%%%%%%%%%%%%%%%%%%
\section{Reduced purities}\label{review}

The distilled purities are derivative quantities of the reduced purities. To define them and understand their significance, it is necessary to  review basic aspects of the reduced purities~\cite{Franco_Reduced}.  Below we present exact expressions for the one and two-body reduced purities, which are the most important and readily applicable reduced purities. The final expression for these reduced purities will lead to expressions for the associated one- and two-body distilled purities. The expressions below generalize the developments in \cite{Franco_Reduced}. As discussed, equations {19 and 21} in Ref. \onlinecite{Franco_Reduced} apply to many-body states with no distinct pairs of Slater determinants that differ by the same particle transition. The equations below overcome this limitation and apply to general electronic states. 

The reduced purities are a hierarchy of measures of decoherence that are based on the well-known hierarchy of $r$-RDM, $\{^{(r)}\h{\Gamma}\}$ ($r=1, 2, \cdots$).  The $r$-RDMs that define the reduced purities are  obtained by tracing out $(N-r)$ electronic degrees of freedom out of $\h{\rho}_e$. Specifically,
\ba\label{rRDM}
^{(r)}{\Gamma}_{\ep_1\ldots \ep_r}^{\ep^{\prime}_1\ldots \ep^{\prime}_r}(t)&=\frac{1}{r!}{\text{Tr}}[{c_{\ep_1}^{\dagger}}{c_{\ep_2}^{\dagger}}\ldots{c_{\ep_r}^{\dagger}}{c_{\ep^{\prime}_r}}\ldots{c_{\ep^{\prime}_2}}{c_{\ep^{\prime}_1}}{\hat{\rho}_{e}(t)}],
\ea
where $c_{\ep_i}^{\dagger}$ (or $c_{\ep_i}$) creates (or annihilates) a fermion in the $i$-th spin-orbital, i.e. $|\ep_i\ra= c_{\ep_i}^{\dagger}\ket{0}$ where $\ket{0}$ is the vacuum state~\cite{Davidson,Fetter}. The  creation and annihilation operators  satisfy the usual fermionic anticommutation relations, $\{c_{\ep_i},c_{\ep_j}^{\dagger}\}=\delta_{\ep_i\ep_j}$ and $\{c_{\ep_i},c_{\ep_j}\}=\{c_{\ep_i}^{\dagger},c_{\ep_j}^{\dagger}\}=0$. The $r$-body reduced purity measures the non-idempotency of $^{(r)}\h{\Gamma}$, as defined in \Eq{eq:rpur}. 
In doing so, it captures coherences in the system that manifest at the $r$-particle level. 

At this point, it is convenient to express $\hat{\rho}_e$ in terms of a basis of Slater determinants 
\ba\label{superposition}
\hat{\rho}_e&=\sum_{n,m} a_{nm}|\Phi_n\rangle\langle \Phi_m|,
\ea
where $\ket{\Phi_n}$ corresponds to a single Slater determinant with integer occupation numbers in a given single particle basis $\{ \ket{\ep_i}\}$. $a_{nn}$ in \Eq{superposition} denote the population of Slater determinant $n$, while $a_{nm}$ refer  to the $S$-coherences between the $n, m$ pair. 

As in \cite{Franco_Reduced}, we define the \emph{order $s_{nm}$ of the $S$-coherence} $a_{nm}$ as the number of single particle transitions required to do a $|\Phi_n\ra \to |\Phi_m\ra$ transition. Specifically,  
\ba\label{order}
s_{nm}=N-\dsp{\sum_{\epsilon} f_n(\epsilon)f_m(\epsilon)}
\ea
where $f_n(\epsilon)=\la \Phi_n|c^{\dagger}_{\ep}c_{\ep}|\Phi_n\ra$ is the distribution function of the state $|\Phi_n\ra$, and adopts values of 0 or 1 based on whether the spin-orbital $|\ep\ra$ is unoccupied or occupied, such that $f_n^2(\epsilon)=f_n(\epsilon)$. The quantity $s_{nm}\in [0, N]$ and takes the value 1 for pairs of states that differ by single excitations, 2 for doubles, etc. 
Note that  $P_r$ in \Eq{eq:rpur} captures $S$-coherences of order $r$ or less.  This is because $S$-coherences between states differing by more than $r$-body transitions do not appear in $^{(r)}\h{\Gamma}$ [\Eq{rRDM}] and are thus not reflected in the $r$-body purity.

To obtain  expressions for $P_1$ and $P_2$ we adopt the expansion in Slater determinants of $\hat{\rho}_e$ in 
\Eq{superposition} and focus on the case where distinct pair of states differ by at most two-particle transitions. Higher-order $S$-coherences do not contribute to $P_1$ and $P_2$, and can be ignored.
Thus, given state $|\Phi_m\ra$ in \Eq{superposition}, all the other states $|\Phi_n\ra$ $(n\neq m)$ in $\h{\rho}_e$ that contribute to the reduced purity are supposed to be at most two-particle transitions away from $|\Phi_m\ra$ such that $s_{nm} \leq 2$. That is,
\ba\label{state}
|\Phi_n\rangle=c^{\dagger}_{\alpha_2}c_{\beta_2}c^{\dagger}_{\alpha_1}c_{\beta_1}|\Phi_m\rangle.
\ea
Here the choice $\alpha_1\neq\beta_1,\alpha_2\neq\beta_2$ is made to guarantee that $|\Phi_n\ra\neq|\Phi_m\ra$.
Also, $\alpha_1\neq\alpha_2, \beta_1\neq\beta_2$ and $c^{\da}_{\beta_1}|\Phi_m\ra=c_{\alpha_1}|\Phi_m\ra=c_{\alpha_2}|\Phi_m\ra=0$ is chosen to prevent $|\Phi_n\ra$ from vanishing.
Note that no particular requirement on the occupation of state $\beta_2$ is adopted to  be able to capture states that differ by a both a single and two-particle transition within this framework. Further note that the labels $\alp_1,\alp_2,\bet_1,\bet_2$ depend on the pair of states $|\Phi_n\ra$ and $|\Phi_m\ra$, and they should always be thought as having an implicit $n,m$ dependence. Such a dependence is not made explicit in the interest of simplicity in the notation.

The detailed calculation of the reduced purities is included in the Appendix. 
The one-body reduced density matrix (1-RDM) associated to $\h{\rho}_e$ is given by
\ba\label{specific1RDM}
\begin{split}
^{(1)}\Gamma_{\ep_1}^{\ep_2}
&=\sum_m a_{mm}\de_{\ep_1\ep_2}f_m(\ep_1)+\dsp{\sum_{\substack{n,m\\n\neq m}}}^\prime a_{nm} f_m(\bet_1)(1-f_m(\alp_1))\\
&\times(1-f_m(\alp_2))\de_{\alp_1\bet_2}\de_{\ep_1\bet_1}\de_{\ep_2\alp_2}.
\end{split}
\ea
Here and throughout the prime in the second sum indicates that only pairs of states that differ by at most two-particle transitions should be considered.
 The resulting expression for the one-body purity is given by
\ba\label{1purity}
\begin{split}
P_1&=\dsp{\sum_{\ep_1,\ep_2}} {}^{(1)}\Gamma_{\ep_1}^{\ep_2} {}^{(1)}\Gamma_{\ep_2}^{\ep_1}
=\sum_{\ep}\left(\sum_{m} a_{mm} f_m(\ep)\right)^2\\
&+\dsp{\sum_{\substack{n,m\\n\neq m}}}^\prime \dsp{\sum_{\substack{p,q\\ p\neq q}}}^\prime a_{nm} a_{pq} A_1\de_{\alp_1\bet_2}\de_{\ga_1\de_2}\de_{\alp_2\de_1}\de_{\bet_1\ga_2},
\end{split}
\ea
where $A_1=f_m(\bet_1)(1-f_m(\alp_1))(1-f_m(\alp_2))f_q(\de_1)(1-f_q(\ga_1))(1-f_q(\ga_2))$. Here we have used the labels $\hat{\rho}_e=\dsp{\sum_{p,q} a_{pq}|\Phi_p\rangle\langle \Phi_q|}$ with $|\Phi_p\ra=c^{\da}_{\ga_2}c_{\de_2}c^{\da}_{\ga_1}c_{\de_1}|\Phi_q\ra$  $(p\ne q)$ to obtain the transpose of 1-RDM, $^{(1)}\Gamma_{\ep_2}^{\ep_1}$, as required to evaluate $P_1$.
The populations of the Slater determinants contribute to the first part of $P_1$, while the $S$-coherences are captured by the second part.
$\de_{\alp_1\bet_2}\, (\de_{\ga_1\de_2})$ in the second term shows that $P_1$ captures $S$-coherences of order 1 between the pair of states $|\Phi_n\ra$ and $|\Phi_m\ra$ $(|\Phi_p\ra$ and $|\Phi_q\ra)$.
Further, distinct pairs of states $n,m$ and $p,q$ contribute to $P_1$ provided they differ by the same one-particle transition as indicated by $\de_{\alp_2\de_1}\de_{\bet_1\ga_2}$.
These contributions due to distinct pairs of states differing by the same one-particle transition were absent in the previously derived Eq.~(19) in Ref.~\cite{Franco_Reduced}. When these states are not present in $\h{\rho}_e$, \Eq{1purity} reduces to Eq.~(19) in Ref.~\cite{Franco_Reduced}.

Similarly, the 2-RDM associated with the many-body density matrix in \Eq{superposition} is
\be\label{specific2RDM}
\begin{split}
^{(2)}\Gamma_{\ep_1,\ep_2}^{\ep_4,\ep_3}
&=\frac{1}{2}\Big[\sum_{m} a_{mm}f_m(\ep_1)f_m(\ep_2)(\de_{\ep_1\ep_4}\de_{\ep_2\ep_3}-\de_{\ep_1\ep_3}\de_{\ep_2\ep_4})\\
&+\dsp{\sum_{\substack{n,m\\n\neq m}}}^\prime a_{nm} f_m(\bet_1)(1-f_m(\alp_1))(1-f_m(\alp_2))\\
&\times f_m(\ep_1)f_m(\ep_2)\left[\de_{\alp_1\bet_2}(\de_{\ep_1\bet_1}(\de_{\ep_2\ep_3}\de_{\alp_2\ep_4}-\de_{\ep_2\ep_4}\de_{\alp_2\ep_3})\right.\\
&\left.\hspace{30mm}-\de_{\ep_2\bet_1}(\de_{\ep_1\ep_3}\de_{\alp_2\ep_4}-\de_{\ep_1\ep_4}\de_{\alp_2\ep_3}))\right.\\
&\left.+(\de_{\ep_1\bet_2}\de_{\ep_2\bet_1}-\de_{\ep_1\bet_1}\de_{\ep_2\bet_2})(\de_{\alp_1\ep_3}\de_{\alp_2\ep_4}-\de_{\alp_2\ep_3}\de_{\alp_1\ep_4})\right]\Big].
\end{split}
\ee
Adopting the same labels as in $P_1$, the final expression for the two-body purity is given by
\be\label{2purity}
\begin{split}
P_2&={\sum_{\ep_1,\ep_2,\ep_3,\ep_4}} {}^{(2)}\Gamma_{\ep_{1},\ep_{2}}^{\ep_{4},\ep_{3}} {}^{(2)}\Gamma_{\ep_{4},\ep_{3}}^{\ep_{1},\ep_{2}}\\
&=\frac{1}{2}\sum_{n,p}a_{nn}a_{pp}\left[\left(\sum_{\ep}f_n(\ep)f_p(\ep)\right)^2-\sum_{\ep}f_n(\ep)f_p(\ep)\right]\\
&+{\sum_{\substack{n,m\\ n\neq m}}}^\prime {\sum_{\substack{p,q\\ p\neq q}}}^\prime a_{nm}a_{pq}A_2\left[\de_{\alp_1\bet_2}\de_{\ga_1\de_2}\de_{\alp_2\de_1}\de_{\bet_1\ga_2}(N-s_{mq})\right.\\
&\left.+A_3(\de_{\bet_1\ga_1}\de_{\bet_2\ga_2}-\de_{\bet_1\ga_2}\de_{\bet_2\ga_1})(\de_{\alp_1\de_1}\de_{\alp_2\de_2}-\de_{\alp_1\de_2}\de_{\alp_2\de_1})\right],
\end{split}
\ee
where $A_2=f_m(\bet_1)(1-f_m(\alp_1))(1-f_m(\alp_2))f_q(\de_1)(1-f_q(\ga_1))(1-f_q(\ga_2))$ and $A_3=f_m(\bet_2)f_q(\de_2)$. 
The diagonal terms of $\h{\rho}_e$ in the single-Slater determinant basis [\Eq{superposition}] appear in the first square bracket of the expression whereas the  $S$-coherences are present in the second square bracket. In addition to population-dependent terms and the  $S$-coherences  already captured by Eq.~(21) in Ref.~\cite{Franco_Reduced}, \Eq{2purity} captures $S$-coherences of order 1 (or 2) between \emph{distinct} pairs of states that differ by the same one-body (or two-body) transition from each other. 
This is imposed by the terms $\de_{\alp_2\de_1}\de_{\bet_1\ga_2}$ and $(\de_{\bet_1\ga_1}\de_{\bet_2\ga_2}-\de_{\bet_1\ga_2}\de_{\bet_2\ga_1})(\de_{\alp_1\de_1}\de_{\alp_2\de_2}-\de_{\alp_1\de_2}\de_{\alp_2\de_1})$ for the one-body and two-body transition respectively.

%%%%%%%%%%%%%%%%%%%%%%%%%%%%%%%%%%%%%%%%%%%%%%%%%%%%%%%%%%%%%%%%%%%%%%
\section{Distilled Purities}\label{distilledpurity}

\subsection{Definition and basic properties}

The reduced purities, while numerically accessible,  are typically difficult to interpret because the non-idempotency of $^{(r)}\h{\Gamma}$ can arise due to decoherence or due  to electronic correlation~\cite{Franco_Reduced}.
In fact, the degree to which $^{(1)}\h{\Gamma}$ deviates from idempotentency is an established measure of electronic correlation in isolated many-electron systems~\cite{Ziesche,Huang}.
Further, even the less demanding goal of attempting to isolate $B$-coherences among general many-particle states at the $r$-RDM level is quite challenging because these $B$-coherences can get all mixed up in $^{(r)}\h{\Gamma}$ when invoking a particular single-particle basis to construct the $r$-RDM. 

One type of $B$-coherences among many-particle states that can, in fact, be isolated at the $r$-RDM level is that among  Slater determinants defined by a given single-particle basis, \emph{i.e.} the $S$-coherences.  As can be seen in  \Eqs{1purity}{2purity}, both the one-body and two-body reduced purity are composed of a term that depends on the populations of the single Slater determinants, while the second term is completely determined by the $S$-coherences.  Note that while the $S$-coherences are not necessarily indicative of the degree of purity of the system, they can provide useful information to interpret the dynamics. The role of the distilled purities introduced below is precisely to extract the contributions of the $S$-coherences to the reduced purities. 

The one-body and two-body distilled purities, $\tilde{P}_1$ and $\tilde{P}_2$, are defined as follows:
\ba
\label{eq:dp1}
\tilde{P}_{1}&=P_1 -\sum_{\epsilon}\left({}^{(1)}\Gamma_{\epsilon}^{\epsilon}\right)^2\\
\label{eq:dp2}
\tilde{P}_{2}&=P_2 -2\sum_{\epsilon_{1},\epsilon_{2}}\left({}^{(2)}\Gamma_{\epsilon_{1}\epsilon_{2}}^{\epsilon_{1}\epsilon_{2}}\right)^2.
\ea
In essence, the second term in this expression \emph{distills} the contributions of the $S$-coherences to $P_r$ by removing the term dependent on the populations of the Slater determinants.  Specifically, the one-body distilled purity is given by
\ba\label{dis1}
\tilde{P}_{1}&=\dsp{{\sum_{\substack{n,m\\n\neq m}}}^\prime {\sum_{\substack{p,q\\ p\neq q}}}^\prime} a_{nm} a_{pq} A_1\de_{\alp_1\bet_2}\de_{\ga_1\de_2}\de_{\alp_2\de_1}\de_{\bet_1\ga_2},
\ea
where $A_1=f_m(\bet_1)(1-f_m(\alp_1))(1-f_m(\alp_2))f_q(\de_1)(1-f_q(\ga_1))(1-f_q(\ga_2))$, and the sums are over pairs of states that differ by at most two particle transitions.  In obtaining \Eq{dis1}, we have used the fact that  $^{(1)}\Gamma_{\epsilon}^{\epsilon}=\sum_n a_{nn}f_n(\ep)$, as can be verified from \Eq{specific1RDM}. In turn, from  \Eqs{specific2RDM}{2purity} it follows that  the two-body purity is given by
\be\label{dis2}
\begin{split}
\tilde{P}_2&={\sum_{n\neq m}}^\prime{\sum_{p\neq q}}^\prime a_{nm}a_{pq}A_2\left[\de_{\alp_1\bet_2}\de_{\ga_1\de_2}\de_{\alp_2\de_1}\de_{\bet_1\ga_2}(N-s_{mq})\right.\\
&\left.+A_3(\de_{\bet_1\ga_1}\de_{\bet_2\ga_2}-\de_{\bet_1\ga_2}\de_{\bet_2\ga_1})(\de_{\alp_1\de_1}\de_{\alp_2\de_2}-\de_{\alp_1\de_2}\de_{\alp_2\de_1})\right],
\end{split}
\ee
where $A_2=f_m(\bet_1)(1-f_m(\alp_1))(1-f_m(\alp_2))f_q(\de_1)(1-f_q(\ga_1))(1-f_q(\ga_2))$ and $A_3=f_m(\bet_2)f_q(\de_2)$. A detailed calculation of the one-body and two-body distilled purities is provided in the Appendix.

The distilled purities provide a succinct way to summarize the $S$-coherences in the system in a particular basis. They are easy to obtain by simple matrix manipulations of the $r$-RDMs as indicated in \Eqs{eq:dp1}{eq:dp2}. 
The one-body distilled purity can capture $S$-coherences between pairs of Slater determinants that differ by at most a one-body transition. In fact, the terms $\de_{\alp_1\bet_2}$ and $\de_{\ga_1\de_2}$  guarantee that this is the case. In turn, $\tilde{P}_2$ captures $S$-coherences of order 1 or 2. 
When distinct pairs of states (i.e. $\{p,q\}\neq\{n,m\}$) appear, they contribute to $\tilde{P}_1$ (or $\tilde{P}_2$) only if they differ by the same one-body (or one and two-body) transition. The distilled purities provide a manifestly basis dependent measure of coherence that succinctly captures the behavior of the off-diagonal elements of the few-body reduced density matrices expressed in a given single-particle basis.

%%%%%%%%%%%%%%%%%%%%%%%%%%%%%%%%%%%%%%%%%%%%%%%%%%%%%%%%%%%%%%%%%%%%%%
\subsection{Limiting values}\label{table}

To aid the interpretation of the dynamics of  $\tilde{P}_1$ and $\tilde{P}_2$ in \Eqs{eq:dp1}{eq:dp2} it is useful to determine a few limiting values. The minimum value for the distilled purities is, of course, $\tilde{P}_r=0$ $(r=1,2)$. This occurs when all $S$-coherences of order $r$ or less are zero, i.e.  $a_{nm} = 0, \, \forall\, n \neq m$. For example, when the state can be described as a single-Slater determinant in the given basis, or when the $S$-coherences in the system are of order greater than $r$. A non-zero distilled purity signals $S$-coherences in the particular basis. The maximum reduced purity is achieved for a pure electronic state that can be described as a single Slater determinant in \emph{some} basis.  In this case, $P_1=N$ and $P_2= N(N+1)/2$~\cite{Franco_Reduced} (when contrasting with the result in \cite{Franco_Reduced}  note that a superposition of single Slater determinant in which all the determinants differ by one particle transitions, $s_{nm}=1$, must also be a single Slater determinant \cite{Zhang14,Ando63}).  Thus, the maximum value for the distilled purity is given by
\be
\begin{split}
\tilde{P}_1^\text{max} & \le P_1^\text{max}  -\sum_{\epsilon}\left({}^{(1)}\Gamma_{\epsilon}^{\epsilon}\right)^2\\
& \le N - \sum_{\ep}\left(\sum_{m} a_{mm} f_m(\ep)\right)^2\\
  & \le  \sum_{m\ne n} a_{mm}a_{nn} s_{nm}\\ 
 \end{split}
\ee
where we have used \Eq{order} and the fact that  $s_{nn}=0$. Note that $s_{nm}$ in the equation above can have any value $s_{nm}\in[0, N]$ because it arises from the contribution of the populations of the Slater determinants. A less restricting inequality can be obtained by taking into account that  $s_{nm}^\text{max}=N$,
\be
\begin{split}
\tilde{P}_1^\text{max} & \le  N \sum_{m\ne n} a_{mm}a_{nn}\\ 
& \le  N (1- \sum_{n} a_{nn}^2). 
 \end{split}
\ee
By the Cauchy-Schwarz inequality [$(\sum_{i}u_i v_i)^2 \le (\sum_{i}u_i^2) (\sum_{i}v_i^2)$], 
\be
 \left(\sum_{n} a_{nn}\right)^2 = 1 \le   \left(\sum_{n}a_{nn}^2\right)  \left(\sum_{m} 1\right) = K \sum_{n}a_{nn}^2
\ee
where $K$ is the total number of determinants that can be constructed in the given basis.  Using this inequality,
\be
\begin{split}
\tilde{P}_1^\text{max} & \le  N \left(1- \frac{1}{K}\right). 
 \end{split}
\ee
By a similar argument,
\be
\begin{split}
\tilde{P}_2^\text{max} & \le   \sum_{n> m} a_{mm}a_{nn} s_{nm} (2N- s_{nm}-1)\\
 & \le \frac{N(N-1)}{2}\left( 1 - \frac{1}{K}\right),
\end{split}
\ee
where the first inequality is significantly more restrictive than the second one.

An increase in the distilled purities from their minimum value of 0 indicates the creation of $S$-coherences in the given basis. An increase (or decrease) in the value of the $S$-coherences will generally lead to an increase (or decrease) in the distilled purities. Exceptions can arise in the case where there are distinct pairs of states that differ by the same single or double particle transition in the superposition as the phase, and not just the magnitude, of these $S$-coherences influence the distilled purities, see \Eqs{eq:ex1}{eq:ex2}.

%%%%%%%%%%%%%%%%%%%%%%%%%%%%%%%%%%%%%%%%%%%%%%%%%%%%%%%%%%%%%%%%%%%%%%
\section{Numerical Examples}\label{results}

\begin{figure*}[htbp]
\includegraphics[width=\textwidth]{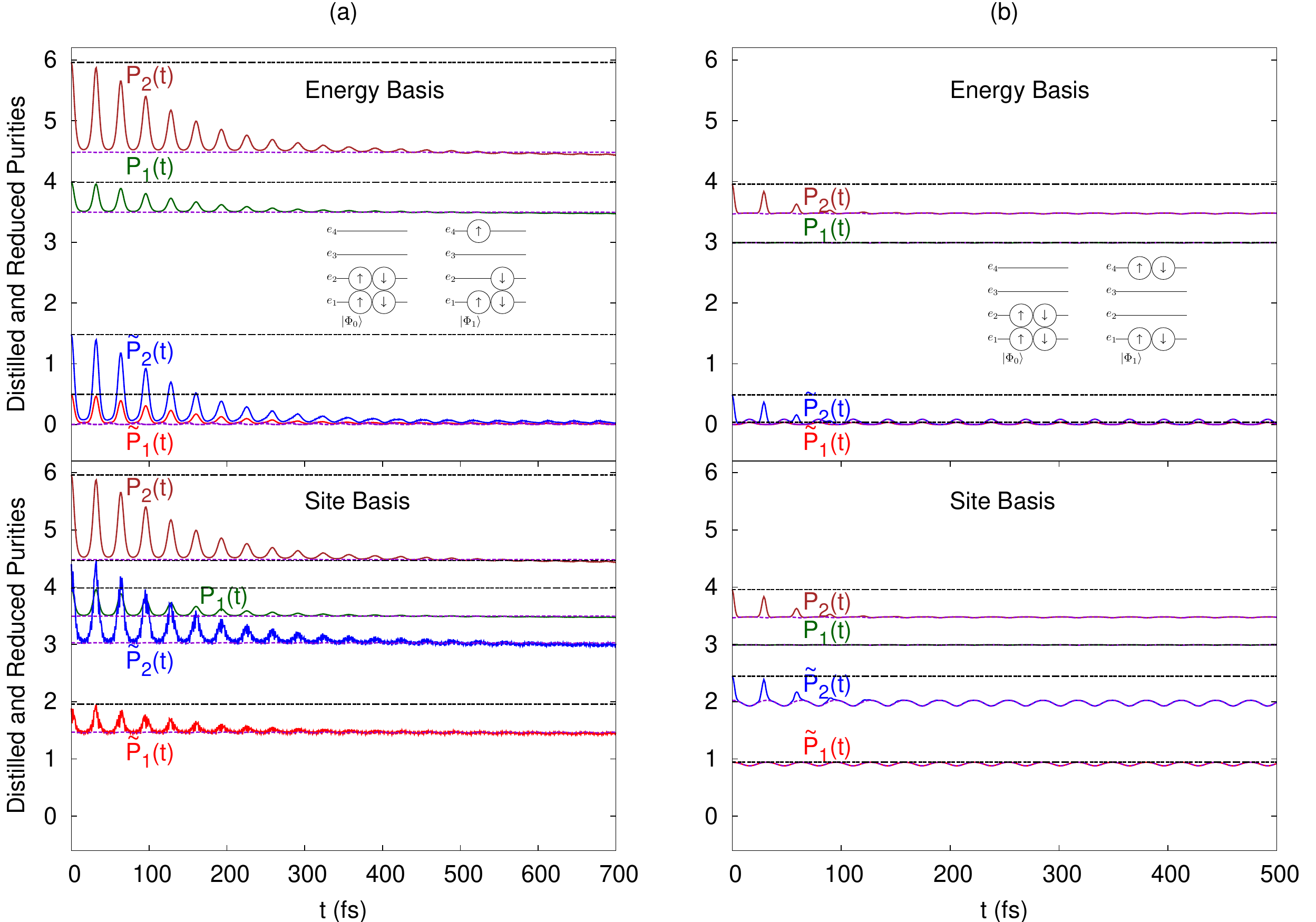} 
\caption{
Distilled (blue, red) and reduced purity (brown, green) in energy and site basis during the vibronic evolution of a neutral  SSH chain with four electrons and clamped ends. The system is initially prepared in a superposition of the electronic ground state $|\Phi_0\ra$ and an excited state $|\Phi_1\ra$ with an initial state $|\Omega(0)\ra=\frac{1}{\sqrt{2}}\left(|\Phi_0\ra+|\Phi_1\ra\right)\otimes|\chi(0)\ra$, where $|\chi(0)\ra$ is the  ground vibrational nuclear state. The occupation of the molecular orbitals in the single Slater determinants $\ket{\Phi_0}$ and $\ket{\Phi_1}$  in each case are shown in the inset. In (a) $|\Phi_0\ra$ and $|\Phi_1\ra$ differ by a one-body transition, while in (b) $|\Phi_0\ra$ and $|\Phi_1\ra$ differ by a two-body transition.  The dashed lines show limiting values for the distilled and reduced purities for: (in black) pure electronic state $\hat{\rho}_e= \ket{\Psi}\bra{\Psi}$ where $\ket{\Psi}=\frac{1}{\sqrt{2}}(\ket{\Phi_0} +  \ket{\Phi_1} )$, and; (in purple) a mixed state of the form  $\hat{\rho}_e = \frac{1}{2}(\ket{\Phi_0}\bra{\Phi_0}+ \ket{\Phi_1}\bra{\Phi_1})$. }
\label{fig1}
\end{figure*}

\begin{figure}[htb]
\includegraphics[width=0.49\textwidth]{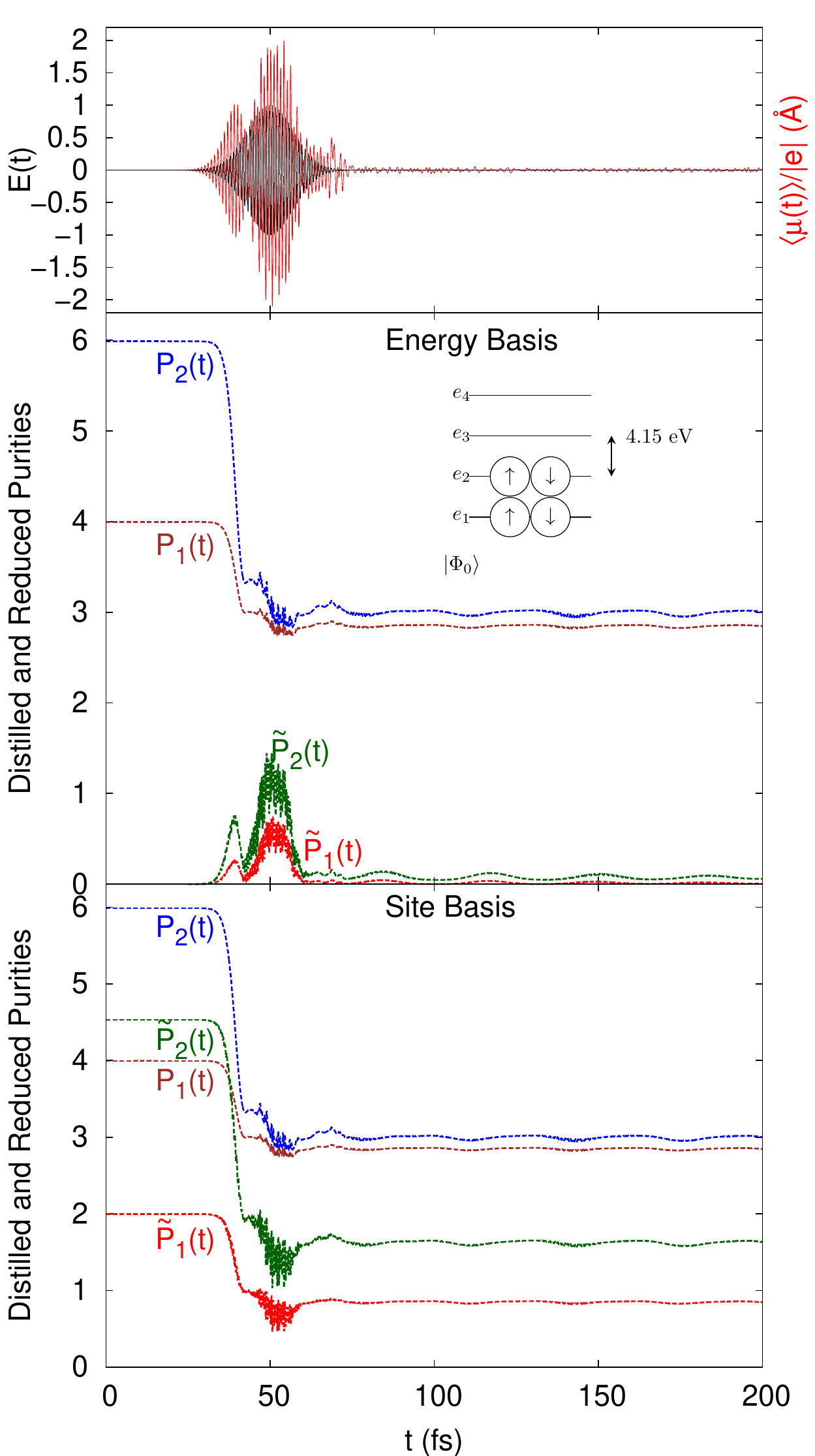}
\caption{Distilled and reduced purity during  resonant photoexcitation of a neutral SSH chain with 4 electrons initially in the ground vibronic state with a 10 fs laser pulse. The laser pulse $E(t)=E_0 e^{-\left(\frac{t-5 t_\omega}{t_\omega}\right)^2}\cos(\omega t)$ (upper panel in black) is chosen to be at resonance with the HOMO-LUMO transition. Here,  $t_w=10$ fs, $E_0= 1.0$ V/\AA\ and $\hbar \omega=4.15\,\text{eV}$. The dipole moment dynamics is shown in red in the upper panel. Notice the growth and  decay of the distilled purities in the energy basis as the field develops, signaling laser-induced $S$-coherences and their eventual decay.}
\label{fig2}
\end{figure}

We now exemplify the behavior of the distilled purities, and contrast it to that of the reduced purities, in the context of a Su-Schrieffer-Heeger (SSH) model for oligoacetylene~\cite{SSH}; a tight-binding model of non-interacting electrons with electron-vibrational couplings. Specifically, we consider the  dynamics of a SSH chain composed of four carbon atoms and four $\pi$ electrons. The four $\pi$ electrons are distributed in four molecular orbitals $|e_i\ra$ of energy $e_i$, leading to 19 possible configurations (without counting spin degeneracies). A detailed discussion of the SSH model and the Ehrenfest mixed quantum-classical technique used to follow the vibronic dynamics has been presented before \cite{Franco_Femtoseconddynamics,Franco_SSH,Franco_Vibret}. Here the electrons are the system of interest, the nuclei are the bath, and the electron-ion coupling is the source of electronic decoherence. 

To test the utility of the distilled purity to inform about dynamical processes in the system, we consider the following exemplifying cases: (i) An initial separable vibronic state in which the electrons are in a superposition  of energy eigenstates, i.e.
\be
\label{eq:initsuper}
\ket{\Omega(0)} = \frac{1}{\sqrt{2}}(\ket{\Phi_0} +  \ket{\Phi_1} )\otimes \ket{\chi_0},\\
\ee
where $\ket{\chi_0}$ is the ground vibrational state associated with the ground electronic state $\ket{\Phi_0}$, and $\ket{\Phi_1}$ is an  excited electronic state. Both $\ket{\Phi_0}$ and $\ket{\Phi_1}$ are taken to be single Slater determinants.  (ii) A chain in the ground vibronic state  subject to a laser pulse that is resonant with a specific electronic transition.  Case (i) is simple to interpret using reduced purities because the population  of the Slater determinants remains approximately constant, and thus the dynamics of the reduced purities reflect the dynamics of the $S$-coherences~\cite{Franco_Reduced}. By contrast, in case (ii) the populations of the Slater determinants involved change in time making it challenging to  separate the dynamics of the $S$-coherences from the dynamics of the populations using the reduced purities. Through (i), we illustrate how the distilled purities reflect the decay of the initial $S$-coherences. Through (ii) we test the ability of the distilled purities to monitor laser-induced $S$-coherences that are obscured by the population dynamics in the reduced-purities. 

To explore the effect of changing the basis, the distilled and reduced purities are computed in the  molecular orbital (energy) $\ket{e_i} = c^{\da}_{e_i}\ket{0}$ and the site $|n\ra=c^{\da}_{n}|0\ra$ basis, where $\ket{0}$ is the vacuum state.  The molecular orbitals are the eigenstates of the single-particle SSH Hamiltonian in the optimal geometry of the chain. In turn, the sites refer to the spatially localized orbitals located at the positions of the carbon atoms in the chain.  Naturally, these two basis are connected via a unitary transformation: $c^{\da}_{n}=\sum_i \la e_i|n\ra c^{\da}_{e_i}$. By applying this transformation to the 1-RDM and 2-RDM, the distilled and reduced purities can be computed in either the site or energy basis using Eqs.~\eqref{eq:rpur}, \eqref{eq:dp1} and \eqref{eq:dp2}. Roughly speaking, the $S$-coherences in site representation signal spatial coherences in the state, while the $S$-coherences in energy representation signal dynamics.

\Fig{fig1} shows the dynamics of the distilled purities for the SSH chain prepared in an initial superposition of the form 
\Eq{eq:initsuper} where $\ket{\Phi_0}$ and $\ket{\Phi_1}$ differ by (a) one or (b) two-particle transitions in the molecular orbital basis as specified in the figure. During the vibronic dynamics of such states, there is evolution of the nuclear wavepacket in the excited state potential energy surface. Such evolution leads to a decay of the nuclear wavepacket overlap $|\la\chi_0(t)|\chi_1(t)\ra|$ associated with the ground $\ket{\Phi_0}$ ($\ket{\chi_0}$) and excited electronic state $\ket{\Phi_1}$ ($\ket{\chi_1}$). Such overlap determines the $S$-coherences between $\ket{\Phi_0}$   and $\ket{\Phi_1}$ and its decay leads to a decay of the purity  of the electronic subsystem (cf. \Eq{Npurity}), and thus to a decay in the reduced purity~\cite{Franco_Reduced}.  As shown in \Fig{fig1}, the distilled purities capture the wavepacket evolution that leads to such decoherence. In both the energy and site basis, the distilled purities display a fast initial decay with recurrences every $\sim30$ fs. These recurrences arise from the time dependence of the overlap of the nuclear wavefunctions in the ground and excited electronic states (see \Eq{Npurity}), and signal the oscillatory motion of the nuclear wavepacket in the excited state potential. Between consecutive recurrences the amplitude of the distilled purity diminishes and eventually reaches an asymptotic value.

Note that in this case the dynamics of the distilled purities closely mimic that of the reduced purities. The reason for this is because in this particular case there are no appreciable changes in the populations of the two Slater determinants involved ($a_{mm}$ in \Eq{1purity}), and thus the dynamics of both quantities is determined by the $S$-coherences.  Nevertheless, while the reduced purities are basis independent,  the value of the distilled purities depend on the single-particle basis employed. In fact, in the energy basis the distilled purities asymptotically go to zero signaling the fact that the $S$-coherences between the Slater determinants constructed using the molecular orbitals basis decays to zero upon time evolution. However, the distilled purities in the site basis do not go to zero indicating that even for the asymptotic state some spatial coherences remain, as is expected for a quantum mechanical system.

Consider now how the distilled purities change with the coherence order. The initial superposition in \Fig{fig1} (a) is of order one, while that of (b) is of order 2.  The superposition in (a) is visible both in $\tilde{P}_1$ and $\tilde{P}_2$, and the fall of $\tilde{P}_2$ is $(N-1)$ times larger than that of $\tilde{P}_1$.  By contrast, in (b) $\tilde{P}_2$ follows the decay of $S$-coherences while $\tilde{P}_1$ remains constant because it cannot distinguish  a coherence of second order from a mixture of such states. At initial time, $\tilde{P}_2$ takes its maximum value that is consistent with the superposition in question and evolves with the vibronic evolution. 

\Fig{fig2} shows the dynamics of the polarization, distilled and the reduced purities during resonant photoexcitation with a 10 fs laser pulse. Such laser creates a superposition of single Slater determinants that is then subject to decoherence due to vibronic couplings. During photoexcitation the one and two body purity decay, as a result of the population of other possible Slater determinants and subsequent decoherence processes after photoexcitation.  Such a decay is mirrored by the distilled purities in the site basis that signal the decay of spatial $S$-coherences that is onset by photoexcitation. Interpreting  the dynamics of the reduced purities is quite challenging as it involves determining all the Slater determinants that participate in the dynamics, their populations and the $S$-coherences among them. By contrast,  the distilled purities in the energy basis clearly show the $S$-coherences that are created by the laser pulse and their eventual decay due to decoherence, as signaled by the growth of the distilled purities and their decay in the energy basis.  The distilled purities in the energy basis attains a maximum at 50 fs when the laser pulse is at its maximum, and follows the dynamics of the polarization as both quantities depend on the $S$-coherences in the energy basis. This example clearly shows how the distilled purities can aid the interpretation of the dynamics of many-body systems by signaling $S$-coherences that are created/destroyed during evolution.

%%%%%%%%%%%%%%%%%%%%%%%%%%%%%%%%%%%%%%%%%%%%%%%%%%%%%%%%%%%%%%%%%%%%%%
\section{Final Remarks}\label{conclude}

\begin{table*}[htbp]
\begin{tabular}{| m{0.18\textwidth} | m{0.27\textwidth} | m{0.5\textwidth} |}
\hline
\textbf{Type} &  \textbf{Definition}  & \textbf{Remarks} \\
\hline
Purity & 
\[
P=\te{Tr}[\h{\rho}_e^2]
\] & 
\begin{itemize}[leftmargin=*]
\item Measures the non-idempotency of $\h{\rho}_e$
\item Well-defined and easy-to-interpret measure of coherence
\item Basis independent 
\item Numerically removed for many-body systems
\end{itemize}
\\
 \hline
Reduced Purity & 
\[
P_r=\te{Tr}[^{(r)}\h{\Gamma}^2]
\] & 
\begin{itemize}[leftmargin=*] 
\item Measures the non-idempotency of $^{(r)}\h{\Gamma}$
\item Difficult to interpret as both decoherence and correlation among electrons lead to non-idempotency of $^{(r)}\h{\Gamma}$
\item Basis independent 
\item Easy to compute
\end{itemize}
\\
\hline
Distilled Purity & 
\[
\tilde{P}_{1}=P_1-\sum_{\epsilon}\left({}^{(1)}\Gamma_{\epsilon}^{\epsilon}\right)^2
\]
\be
\begin{split}
\tilde{P}_{2}=P_2
-2\sum_{\epsilon_{1},\epsilon_{2}} \left({}^{(2)}\Gamma_{\epsilon_{1}\epsilon_{2}}^{\epsilon_{1}\epsilon_{2}}\right)^2\nn
\end{split}
\ee
&
\begin{itemize}[leftmargin=*]
\item Summarizes $S$-coherences (off-diagonal elements among Slater determinants defined by a given single particle basis)
\item Useful and easy to interpret, but not necessarily informative of state purity 
\item Basis dependent 
\item Easy to compute
\end{itemize}
\\
\hline
\end{tabular}
\caption{Basic features of the different ways to quantify electronic decoherence  in many-body systems.}
\label{tab: table2}
\end{table*}

The basic features of the three measures of electronic decoherence discussed in this paper- purity, reduced purity and distilled purity are summarized in Table~\ref{tab: table2}. The purity is a well-defined basis-independent measure of coherence that directly signals the extent to which the electronic subsystem is described as a mixed state. 
Whenever possible, this is our preferred quantity to interpret decoherence. However, to obtain it one needs the $N$-particle electronic density matrix which is generally inaccessible, making the purity often impractical to measure electronic decoherence in  many body systems.

The reduced purities introduced in Ref.~\cite{Franco_Reduced} measure the non-idempotency of the $r$-RDMs. These quantities are basis independent and  accessible from simulations that propagate the 1-RDM and 2-RDM directly. For non-interacting electronic systems the decay of the reduced purity  directly signals coherence loss.
Nevertheless, in the general case where both electron-nuclear and electron-electron interactions play a role in the dynamics, the decay of the reduced purity can come from electronic correlation or from decoherence. Since these two effects are challenging to separate at the $r$-RDM level,  the reduced purities are of limited applicability as a measure of electronic coherence or correlation. 

As a practical alternative, here we have introduced the one- and two-body distilled purities in \Eqs{eq:dp1}{eq:dp2} as a tool to interpret the dynamics of many-body systems in the presence of decoherence. The distilled purities are derivative quantities of the reduced purities that distill the contributions of the $S$-coherences to the reduced purities. That is, the distilled purities summarize the $S$-coherences among $N$-particle single Slater determinant states with integer occupations as defined by  a given single particle basis. In this analysis, we have  derived exact expression for the one-body and two-body distilled purities for general electronic states. For this, we generalized  the expressions for the one- and two-body reduced purities in Ref.~\cite{Franco_Reduced}, by capturing possible contributions coming from two distinct pairs of states that differ by the same one- or two-particle transition. 

The distilled purities are manifestly basis-dependent quantities that are useful in interpreting the dynamics of many-body systems. As an example, the distilled purities were shown to be able to signal $S$-coherences that are generated during resonant photoexcitation of a model molecule, which are obscure in the reduced purities. Further, since the $r$-body distilled purities can capture $S$-coherences of order $r$ or less, investigating the behavior of the distilled purities of different orders can aid the interpretation of the many-body dynamics. In spite of these advantages, the distilled purity is not simply related to the $N$-body purity of the system, and thus it is not indicative of the degree of coherence of the system.  For example, a pure electronic state that can be described as a single Slater determinant in a given basis will have a distilled purity of zero in such basis. This limitation is shared with other basis-dependent measures of coherence. For instance, in the energy eigenbasis a ground state molecule in a pure state will have no off-diagonal elements in the density matrix  and thus zero $B$-coherences, even when it is in a pure state. Albeit not necessarily indicative of whether there is actual decoherence in the system, these quantities are useful in analyzing the  quantum dynamics of many-body systems in a situation where the purity is an inaccessible quantity.

%%%%%%%%%%%%%%%%%%%%%%%%%%%%%%%%%%%%%%%%%%%%%%%%%%%%%%%%%%%%%%%%%%%%%%
\begin{acknowledgments}
This material is based upon work supported by the National Science Foundation under CHE - 1553939. 
\end{acknowledgments}

%%%%%%%%%%%%%%%%%%%%%%%%%%%%%%%%%%%%%%%%%%%%%%%%%%%%%%%%%%%%%%%%%%%%%%
\appendix*
\section{Derivation of the reduced and distilled purities}
\label{appen}

Below we derive the one- and two-body reduced and distilled purities (\Eqs{1purity}{2purity} and \Eqs{dis1}{dis2}) for the 
general electronic state in \Eq{superposition}.

%%%%%%%%%%%%%%%%%%%%%%%%%%%%%%%%%%%%%%%%%%%%%%%%%%%%%%%%%%%%%%%%%%%%%%
\subsection{One-body reduced and distilled purity}

%%%%%%%%%%%%%%%%%%%%%%%%%%%%%%%%%%%%%%%%%%%%%%%%%%%%%%%%%%%%%%%%%%%%%%
\subsubsection{1-RDM}

The 1-RDM for a general electronic state of the form in \Eq{superposition} is given by 
\ba\label{1RDM}
\begin{split}
^{(1)}\Gamma_{\ep_1}^{\ep_2}&=\text{Tr}[c_{\ep_1}^{\da} c_{\ep_2} \hat{\rho}_{e}]\\
&=\dsp{\sum_{n} a_{nn} \la\Phi_n|c^{\da}_{\ep_1}c_{\ep_2}|\Phi_n\ra}
+\dsp{\sum_{\substack{n,m\\n\neq m}}} a_{nm} \la\Phi_m|c_{\ep_1}^{\da} c_{\ep_2}|\Phi_n\ra\\
&=\dsp{\sum_{n} a_{nn} \la\Phi_n|c^{\da}_{\ep_1}c_{\ep_2}|\Phi_n\ra}\\
&+\dsp{{\sum_{\substack{n,m\\n\neq m}}}^\prime} a_{nm}\la\Phi_m|c_{\ep_1}^{\da} c_{\ep_2}c^{\da}_{\alp_2}c_{\bet_2}c^{\da}_{\alp_1}c_{\bet_1}|\Phi_m\ra,
\end{split}
\ea
where we have used  \Eqs{rRDM}{state}, and where the prime indicates that the sum goes over pairs of states that differ by at most two-particle transitions. Note that in the last summation only those states that differ by a single particle transition contribute, as pairs with coherences of higher order are not visible in the 1-RDM. As mentioned in the text, the labels  $\alp_1, \alp_2, \bet_1, \bet_2$ have an implicit dependence on $n$ and $m$.  The second term can be developed further by first taking the creation and annihilation operators into normal ordering and  then employing the restrictions on the  $\alp_1, \alp_2, \bet_1, \bet_2$ detailed under \Eq{state} in Sec.~\ref{review}:
\begin{equation}
\begin{split}
&\la\Phi_m|c_{\ep_1}^{\da} c_{\ep_2}c^{\da}_{\alp_2}c_{\bet_2}c^{\da}_{\alp_1}c_{\bet_1}|\Phi_m\ra\\
&=f_m(\bet_1)(1-f_m(\alp_1))(1-f_m(\alp_2))\\
&\times \la\Phi_m|\left(\de_{\ep_2\alp_2}\de_{\alp_1\bet_2}c^{\da}_{\ep_1}c_{\bet_1}+\de_{\alp_1\bet_2}c^{\da}_{\ep_1}c^{\da}_{\alp_2}c_{\bet_1}c_{\ep_2}\right.\\
&\left.+\de_{\alp_2\ep_2}c^{\da}_{\ep_1}c^{\da}_{\alp_1}c_{\bet_1}c_{\bet_2}+\de_{\alp_1\ep_2}c^{\da}_{\ep_1}c^{\da}_{\alp_2}c_{\bet_2}c_{\bet_1}\right.\\
&\left.+c^{\da}_{\ep_1}c^{\da}_{\alp_2}c^{\da}_{\alp_1}c_{\bet_1}c_{\bet_2}c_{\ep_2}\right)|\Phi_m \ra\\
&=\de_{\alp_1\bet_2}\de_{\ep_1\bet_1}\de_{\ep_2\alp_2}f_m(\bet_1)(1-f_m(\alp_1))(1-f_m(\alp_2)).
\label{eq:interm}
\end{split}
\end{equation}
 By inserting \Eq{eq:interm} into \Eq{1RDM}, it then follows that
\ba
\label{appen:specific1RDM}
^{(1)}\Gamma_{\ep_1}^{\ep_2}&=\sum_m a_{mm}X+\dsp{\sum_{\substack{n,m\\n\neq m}}}^\prime a_{nm} Y,
\ea
where $X=\de_{\ep_1\ep_2}f_m(\ep_1), Y=f_m(\bet_1)(1-f_m(\alp_1))(1-f_m(\alp_2))\de_{\alp_1\bet_2}\de_{\ep_1\bet_1}\de_{\ep_2\alp_2}$ which is Eq. (9) in the main text. For obtaining the one body purity, it is also useful to express the transpose of the 1-RDM in \Eq{1RDM} with a different set of labels $p$ and $q$  ($\hat{\rho}_e=\dsp{\sum_{p,q} a_{pq}|\Phi_p\rangle\langle \Phi_q|}$ and $|\Phi_p\ra=c^{\da}_{\ga_2}c_{\de_2}c^{\da}_{\ga_1}c_{\de_1}|\Phi_q\ra$) as follows:
\ba\label{specific1RDMb}
^{(1)}\Gamma_{\ep_2}^{\ep_1}&=\dsp{\sum_q} a_{qq} Z+\dsp{\sum_{\substack{p,q\\p\neq q}}}^\prime a_{pq} W,
\ea
where $Z=\de_{\ep_2\ep_1}f_q(\ep_2), W=f_q(\de_1)(1-f_q(\ga_1))(1-f_q(\ga_2))\de_{\ga_1\de_2}\de_{\ep_2\de_1}\de_{\ep_1\ga_2}$.
Note that the labels $\ga_1, \ga_2, \de_1, \de_2$ depend on $p,q$ implicitly.
The expressions in \Eqs{appen:specific1RDM}{specific1RDMb} for 1-RDM are now employed to find $P_1$.

%%%%%%%%%%%%%%%%%%%%%%%%%%%%%%%%%%%%%%%%%%%%%%%%%%%%%%%%%%%%%%%%%%%%%%
\subsubsection{One-body reduced purity}

The one-body reduced purity is given by
\ba\label{sim}
\begin{split}
P_1&=\te{Tr}[^{(1)}\h{\Gamma}^2]=\dsp{\sum_{\ep_1,\ep_2}} {}^{(1)}\Gamma_{\ep_1}^{\ep_2} {}^{(1)}\Gamma_{\ep_2}^{\ep_1}\\
&=\dsp{\sum_{\ep_1,\ep_2}}\left(\sum_m a_{mm}X+\dsp{\sum_{\substack{n,m\\n\neq m}}}^\prime a_{nm} Y\right)\left(\dsp{\sum_q} a_{qq} Z+\dsp{\sum_{\substack{p,q\\p\neq q}}}^\prime a_{pq} W\right)\\
&=\dsp{\sum_{\ep_1,\ep_2}}\left[\sum_{m,q} a_{mm}a_{qq} XZ +\dsp{{\sum_{\substack{n,m\\n\neq m}}}^\prime \sum_q} a_{nm} a_{qq} YZ\right.\\
&\left.+\dsp{\sum_m {\sum_{\substack{p,q\\p\neq q}}}}^\prime a_{mm} a_{pq} XW+\dsp{{\sum_{\substack{n,m\\n\neq m}}}^\prime  {\sum_{\substack{p,q\\p\neq q}}}}^\prime a_{nm} a_{pq} YW\right],
\end{split}
\ea
where 
\be
\begin{split}
XZ&=\de_{\ep_1\ep_2}f_m(\ep_1)f_q(\ep_2),\\
YZ&=\de_{\ep_1\ep_2}\de_{\alp_1\bet_2}\de_{\ep_1\bet_1}\de_{\ep_2\alp_2}f_q(\ep_2)f_m(\bet_1)\\
&\times(1-f_m(\alp_1))(1-f_m(\alp_2))=0,\\
XW&=\de_{\ep_1\ep_2}\de_{\ga_1\de_2}\de_{\ep_2\de_1}\de_{\ep_1\ga_2}f_m(\ep_1)f_q(\de_1)\\
&\times(1-f_q(\ga_1))(1-f_q(\ga_2))=0,\\
YW&=A_1\de_{\alp_1\bet_2}\de_{\ep_1\bet_1}\de_{\ep_2\alp_2}\de_{\ga_1\de_2}\de_{\ep_2\de_1}\de_{\ep_1\ga_2}
\end{split}
\ee
and $A_1=f_m(\bet_1)(1-f_m(\alp_1))(1-f_m(\alp_2)) f_q(\de_1)(1-f_q(\ga_1))(1-f_q(\ga_2))$. By removing the terms that vanish, and simplifying one obtains a final expression for the one-body purity [\Eq{1purity}]
\be
\label{appen:1purity}
\begin{split}
P_1&=\sum_{\ep}\left(\sum_{m} a_{mm} f_m(\ep)\right)^2\\
&+\dsp{\sum_{\substack{n,m\\n\neq m}}}^\prime \dsp{\sum_{\substack{p,q\\p\neq q}}}^\prime a_{nm} a_{pq} A_1\de_{\alp_1\bet_2}\de_{\ga_1\de_2}\de_{\alp_2\de_1}\de_{\bet_1\ga_2}.
\end{split}
\ee
This equation can be simplified further by noticing that $\de_{\alp_1\bet_2}(\de_{\ga_1\de_2})$ implies that the pair $n,m\,(p,q)$ are connected by a one-body transition. 
Thus
\ba\label{1grp}
\begin{split}
P_1&=\sum_{\ep}\left(\sum_{m} a_{mm} f_m(\ep)\right)^2\\
&+\dsp{\sum_{\substack{n,m\\n\neq m}}}\dsp{\sum_{\substack{p,q\\p\neq q}}} a_{nm} a_{pq} 
A_1\de_{s_{nm},1}\de_{s_{pq},1}\de_{\alp_2\de_1}\de_{\bet_1\ga_2}.
\end{split}
\ea

%%%%%%%%%%%%%%%%%%%%%%%%%%%%%%%%%%%%%%%%%%%%%%%%%%%%%%%%%%%%%%%%%%%%%%
\subsubsection{One-body distilled purity}

To calculate the one-body distilled purity in \Eq{eq:dp1}, it is necessary to obtain the square of the diagonal element of $^{(1)}\hat{\Gamma}$. From \Eq{appen:specific1RDM}, 
\be
\sum_{\epsilon}\left({}^{(1)}\Gamma_{\epsilon}^{\epsilon}\right)^2 = \sum_{\ep}\left(\sum_{m} a_{mm} f_m(\ep)\right)^2.
\ee
This is the exactly same as the first term in \Eq{1grp}. Thus the one-body distilled purity in \Eq{eq:dp1} can be simplified to 
\be
\begin{split}
\tilde{P}_1&=\dsp{\sum_{\substack{n,m\\n\neq m}}}\dsp{\sum_{\substack{p,q\\p\neq q}}} a_{nm} a_{pq} 
A_1\de_{s_{nm},1}\de_{s_{pq},1}\de_{\alp_2\de_1}\de_{\bet_1\ga_2},
\end{split}
\ee
where $A_1=f_m(\bet_1)(1-f_m(\alp_1))(1-f_m(\alp_2)) f_q(\de_1)(1-f_q(\ga_1))(1-f_q(\ga_2))$. The equation above is equivalent to \Eq{dis1}.

%%%%%%%%%%%%%%%%%%%%%%%%%%%%%%%%%%%%%%%%%%%%%%%%%%%%%%%%%%%%%%%%%%%%%%
\subsubsection{Example}

As a simple example, consider $P_1$ for a 2-particle system with $\hat{\rho}_e=\sum_{n,m=1}^4 a_{nm}|\Phi_n\rangle\langle \Phi_m|$ where $|\Phi_1\ra=c^{\da}_1c^{\da}_2|0\ra,|\Phi_2\ra=c^{\da}_3c^{\da}_4|0\ra,|\Phi_3\ra=c^{\da}_1c^{\da}_4|0\ra,|\Phi_4\ra=c^{\da}_2c^{\da}_3|0\ra$. In this case, \Eq{1grp} yields
\ba
\begin{split}
P_1&=(a_{11}+a_{33})^2+(a_{11}+a_{44})^2+(a_{22}+a_{44})^2\\
&+(a_{22}+a_{33})^2+2\left(|a_{13}|^2+|a_{14}|^2+|a_{23}|^2+|a_{24}|^2\right)\\
&-2\left(a_{14}a_{23}+a_{24}a_{13}+a_{13}^*a_{24}^*+a_{23}^*a_{14}^*\right).
\end{split}
\ea
The associated distilled purity is then:
\ba\label{eq:ex1}
\begin{split}
\tilde{P}_1&= 2\left(|a_{13}|^2+|a_{14}|^2+|a_{23}|^2+|a_{24}|^2\right)\\
&-2\left(a_{14}a_{23}+a_{24}a_{13}+a_{13}^*a_{24}^*+a_{23}^*a_{14}^*\right).
\end{split}
\ea
Notice that the reduced purity is composed of a part that depends on the populations of the Slater determinants and another one on the $S$-coherences. The distilled purities extract the contributions due to the $S$-coherences. The $S$-coherences between each pair of states that differ by a one body transition contribute to $P_1$ and $\tilde{P}_1$.  In addition, there are additional contributions in $P_1$ that arise when two distinct pair of states differ by the same one-particle transition.
For example, the term $a_{14}a_{23}$ appears in $P_1$ because both pairs of states ($\ket{\Phi_1}$ and $\ket{\Phi_4}$, and $\ket{\Phi_2}$ and $\ket{\Phi_3}$) differ by the same one-body transition as $|\Phi_2\ra=c^{\da}_3 c_1 |\Phi_3\ra$ and $|\Phi_4\ra=-c^{\da}_3c_1 |\Phi_1\ra$.
The negative sign in the expression arises from the ordering of the states.

%%%%%%%%%%%%%%%%%%%%%%%%%%%%%%%%%%%%%%%%%%%%%%%%%%%%%%%%%%%%%%%%%%%%%%
\subsection{Two-body reduced and distilled purity}

%%%%%%%%%%%%%%%%%%%%%%%%%%%%%%%%%%%%%%%%%%%%%%%%%%%%%%%%%%%%%%%%%%%%%%
\subsubsection{2-RDM }

The 2-RDM for the state in \Eq{superposition} is given by
\be
\label{eq:2RDM}
\begin{split}
^{(2)}\Gamma_{\ep_1,\ep_2}^{\ep_4,\ep_3}&=\frac{1}{2}\text{Tr}[c_{\ep_1}^{\da} c^{\da}_{\ep_2} c_{\ep_3}c_{\ep_4}\hat{\rho}_{e}]\\
&=\frac{1}{2}\dsp{\sum_{m}} a_{mm} \la\Phi_m|c_{\ep_1}^{\da} c^{\da}_{\ep_2} c_{\ep_3}c_{\ep_4}|\Phi_m\ra\\
&+\frac{1}{2}\dsp{\sum_{\substack{n,m\\n\neq m}}}^\prime a_{nm} \la\Phi_m|c_{\ep_1}^{\da} c^{\da}_{\ep_2} c_{\ep_3}c_{\ep_4}c^{\da}_{\alp_2}c_{\bet_2}c^{\da}_{\alp_1}c_{\bet_1}|\Phi_m\ra.
\end{split}
\ee
where we have used  \Eqs{rRDM}{state}. Note that in the last summation only those states that differ by one or two particle transitions contribute, as pairs with coherences of higher order are not visible in the 2-RDM. This summation can be developed further by adopting normal ordering and imposing  the restrictions on $\alp_1, \alp_2, \bet_1$, and  $\bet_2$:
\ba\label{appen:specific2RDMa}
\begin{split}
&\la\Phi_m|c_{\ep_1}^{\da} c^{\da}_{\ep_2} c_{\ep_3}c_{\ep_4}c^{\da}_{\alp_2}c_{\bet_2}c^{\da}_{\alp_1}c_{\bet_1}|\Phi_m\ra\\
&=f_m(\bet_1)(1-f_m(\alp_1))(1-f_m(\alp_2))\\
&\times \la\Phi_m|\left(\de_{\alp_2\ep_4}\de_{\alp_1\bet_2}c^{\da}_{\ep_1}c^{\da}_{\ep_2}c_{\ep_3}c_{\bet_1}
-\de_{\alp_2\ep_4}\de_{\alp_1\ep_3}c^{\da}_{\ep_1}c^{\da}_{\ep_2}c_{\bet_2}c_{\bet_1}\right.\\
&\left.-\de_{\alp_2\ep_4}c^{\da}_{\ep_1}c^{\da}_{\ep_2}c^{\da}_{\alp_1}c_{\bet_2}c_{\ep_3}c_{\bet_1}
+\de_{\alp_1\bet_2}\de_{\alp_2\ep_3}c^{\da}_{\ep_1}c^{\da}_{\ep_2}c_{\bet_1}c_{\ep_4}\right.\\
&\left.-\de_{\alp_1\bet_2}c^{\da}_{\ep_1}c^{\da}_{\ep_2}c^{\da}_{\alp_2}c_{\ep_4}c_{\ep_3}c_{\bet_1}
-\de_{\alp_1\ep_4}\de_{\alp_2\ep_3}c^{\da}_{\ep_1}c^{\da}_{\ep_2}c_{\bet_1}c_{\bet_2}\right.\\
&\left.+\de_{\alp_1\ep_4}c^{\da}_{\ep_1}c^{\da}_{\ep_2}c^{\da}_{\alp_2}c_{\ep_3}c_{\bet_1}c_{\bet_2}
-\de_{\alp_2\ep_3}c^{\da}_{\ep_1}c^{\da}_{\ep_2}c^{\da}_{\alp_1}c_{\bet_2}c_{\bet_1}c_{\ep_4}\right.\\
&\left.+\de_{\alp_1\ep_3}c^{\da}_{\ep_1}c^{\da}_{\ep_2}c^{\da}_{\alp_2}c_{\ep_4}c_{\bet_2}c_{\bet_1}
+c^{\da}_{\ep_1}c^{\da}_{\ep_2}c^{\da}_{\alp_2}c^{\da}_{\alp_1}c_{\bet_2}c_{\ep_4}c_{\ep_3}c_{\bet_1}\right)|\Phi_m\ra\\
&=f_m(\bet_1)(1-f_m(\alp_1))(1-f_m(\alp_2))f_m(\ep_1)f_m(\ep_2)\\
&\times\left[\de_{\alp_1\bet_2}(\de_{\ep_1\bet_1}(\de_{\ep_2\ep_3}\de_{\alp_2\ep_4}-\de_{\ep_2\ep_4}\de_{\alp_2\ep_3})\right.\\
&\left.
-\de_{\ep_2\bet_1}(\de_{\ep_1\ep_3}\de_{\alp_2\ep_4}-\de_{\ep_1\ep_4}\de_{\alp_2\ep_3}))\right.\\
&\left.+(\de_{\ep_1\bet_2}\de_{\ep_2\bet_1}-\de_{\ep_1\bet_1}\de_{\ep_2\bet_2})(\de_{\alp_1\ep_3}\de_{\alp_2\ep_4}-\de_{\alp_2\ep_3}\de_{\alp_1\ep_4})\right].
\end{split}
\ea
Inserting this expression into \Eq{eq:2RDM} we obtain a final expression for the 2-RDM [\Eq{specific2RDM}]:
\be
\label{appen:specific2RDM}
\begin{split}
^{(2)}\Gamma_{\ep_1,\ep_2}^{\ep_4,\ep_3}&=\frac{1}{2}\Big[\sum_{n} a_{nn}f_n(\ep_1)f_n(\ep_2)(\de_{\ep_1\ep_4}\de_{\ep_2\ep_3}-\de_{\ep_1\ep_3}\de_{\ep_2\ep_4})\\
&+\dsp{\sum_{\substack{n,m\\n\neq m}}}^\prime a_{nm} f_m(\bet_1)(1-f_m(\alp_1))(1-f_m(\alp_2))f_m(\ep_1)f_m(\ep_2)\\
&\times \left[\de_{\alp_1\bet_2}(\de_{\ep_1\bet_1}(\de_{\ep_2\ep_3}\de_{\alp_2\ep_4}-\de_{\ep_2\ep_4}\de_{\alp_2\ep_3})\right.\\
&\left.-\de_{\ep_2\bet_1}(\de_{\ep_1\ep_3}\de_{\alp_2\ep_4}-\de_{\ep_1\ep_4}\de_{\alp_2\ep_3}))\right.\\
&\left.+(\de_{\ep_1\bet_2}\de_{\ep_2\bet_1}-\de_{\ep_1\bet_1}\de_{\ep_2\bet_2})(\de_{\alp_1\ep_3}\de_{\alp_2\ep_4}-\de_{\alp_2\ep_3}\de_{\alp_1\ep_4})\right]\Big].
\end{split}
\ee
To calculate the purity it is also useful to obtain an expression for the transpose of \Eq{appen:specific2RDM} with alternative indexes. Specifically, we employ $\hat{\rho}_e=\displaystyle{\sum_{p,q} a_{pq}|\Phi_p\rangle\langle \Phi_q|}$ and $|\Phi_p\ra=c^{\da}_{\ga_2}c_{\de_2}c^{\da}_{\ga_1}c_{\de_1}|\Phi_q\ra$). In  this case
\be
\label{specific2RDMb}
\begin{split}
^{(2)}\Gamma_{\ep_4,\ep_3}^{\ep_1,\ep_2}&=\frac{1}{2}\Big[\dsp{\sum_{p}} a_{pp}f_p(\ep_4)f_p(\ep_3)(\de_{\ep_4\ep_1}\de_{\ep_3\ep_2}-\de_{\ep_4\ep_2}\de_{\ep_3\ep_1})\\
&+\dsp{\sum_{\substack{p,q\\p\neq q}}}^\prime a_{pq} f_q(\de_1)(1-f_q(\ga_1))(1-f_q(\ga_2))\\
&\times f_q(\ep_4)f_q(\ep_3)\left[\de_{\ga_1\de_2}(\de_{\ep_4\de_1}(\de_{\ep_3\ep_2}\de_{\ga_2\ep_1}-\de_{\ep_3\ep_1}\de_{\ga_2\ep_2})\right.\\
&\left.-\de_{\ep_3\de_1}(\de_{\ep_4\ep_2}\de_{\ga_2\ep_1}-\de_{\ep_4\ep_1}\de_{\ga_2\ep_2}))\right.\\
&\left.+(\de_{\ep_4\de_2}\de_{\ep_3\de_1}-\de_{\ep_4\de_1}\de_{\ep_3\de_2})(\de_{\ga_1\ep_2}\de_{\ga_2\ep_1}-\de_{\ga_2\ep_2}\de_{\ga_1\ep_1})\right]\Big].
\end{split}
\ee
The expressions in \Eqs{appen:specific2RDM}{specific2RDMb} are now employed to find $P_2$.

%%%%%%%%%%%%%%%%%%%%%%%%%%%%%%%%%%%%%%%%%%%%%%%%%%%%%%%%%%%%%%%%%%%%%%
\subsubsection{Two-body reduced purity}

The two-body reduced purity is given by
\ba\label{2pur}
\begin{split}
P_2&=\sum_{\ep_1,\ep_2,\ep_3,\ep_4}{}^{(2)}\Gamma_{\ep_1,\ep_2}^{\ep_4,\ep_3}{}^{(2)}\Gamma^{\ep_1,\ep_2}_{\ep_4,\ep_3}\\
&=\frac{1}{4}\left(\sum_n a_{nn}A+{\sum_{\substack{n,m\\n\neq m}}}^\prime a_{nm}B\right)\left(\sum_p a_{pp}C+{\sum_{\substack{p,q\\p\neq q}}}^\prime a_{pq}D\right)\\
&=\frac{1}{4}\left[\sum_{n,p} a_{nn}a_{pp} A C+{\sum_{\substack{n,m\\n\neq m}}}^\prime \sum_p a_{nm}a_{pp} BC\right.\\
&\left.+\sum_n{\sum_{\substack{p,q\\p\neq q}}}^\prime a_{nn}a_{pq} AD+{\sum_{\substack{n,m\\n\neq m}}}^\prime {\sum_{\substack{p,q\\p\neq q}}}^\prime a_{nm}a_{pq}BD\right],
\end{split}
\ea
where
\ba
AC&=\sum_{\ep_1,\ep_2,\ep_3,\ep_4} f_n(\ep_1)f_n(\ep_2)f_p(\ep_3)f_p(\ep_4)(\de_{\ep_1\ep_4}\de_{\ep_2\ep_3}-\de_{\ep_1\ep_3}\de_{\ep_2\ep_4})^2\nn \\
&=2\sum_{\ep_1,\ep_2}f_n(\ep_1)f_p(\ep_1)f_n(\ep_2)f_p(\ep_2)-2\sum_{\ep_1}f_n(\ep_1)f_p(\ep_1),\nn \\
BD&=4 A_2\Big[\de_{\alp_1\bet_2}\de_{\ga_1\de_2}\de_{\alp_2\de_1}\de_{\ga_2\bet_1}(N-s_{mq})\nn \\
&+A_3(\de_{\bet_1\ga_1}\de_{\bet_2\ga_2}-\de_{\bet_1\ga_2}\de_{\bet_2\ga_1})(\de_{\alp_1\de_1}\de_{\alp_2\de_2}-\de_{\alp_1\de_2}\de_{\alp_2\de_1})\Big]\nn,
\ea
with $A_2=f_m(\bet_1)f_q(\de_1)(1-f_m(\alp_1))(1-f_m(\alp_2))(1-f_q(\ga_1))(1-f_q(\ga_2))$ and $A_3=f_m(\bet_2)f_q(\de_2)$. The terms $BC$ and $AD$ vanish after simplification because of the constraints on $\alp_1, \alp_2, \bet_1, \bet_2, \ga_1, \ga_2, \de_1, \de_2$. Thus, the final expression for the two-body reduced purity is [\Eq{2purity}]:
\ba\label{appen:2purity}
\begin{split}
P_2&=\frac{1}{2}\sum_{n,p}a_{nn}a_{pp}\left[\left(\sum_{\ep}f_n(\ep)f_p(\ep)\right)^2-\sum_{\ep}f_n(\ep)f_p(\ep)\right]\\
&+{\sum_{\substack{n,m\\n\neq m}}}^\prime {\sum_{\substack{p,q\\p\neq q}}}^\prime a_{nm}a_{pq}A_2\left[\de_{s_{nm},1}\de_{s_{pq,1}}\de_{\alp_2\de_1}\de_{\bet_1\ga_2}(N-s_{mq})\right.\\
&\left.+A_3(\de_{\bet_1\ga_1}\de_{\bet_2\ga_2}-\de_{\bet_1\ga_2}\de_{\bet_2\ga_1})(\de_{\alp_1\de_1}\de_{\alp_2\de_2}-\de_{\alp_1\de_2}\de_{\alp_2\de_1})\right].
\end{split}
\ea

%%%%%%%%%%%%%%%%%%%%%%%%%%%%%%%%%%%%%%%%%%%%%%%%%%%%%%%%%%%%%%%%%%%%%%
\subsubsection{Two-body distilled purity}

To calculate the distilled purity using \Eq{eq:dp2}, it is necessary to first determine the square of the  diagonal elements of $^{(2)}\hat{\Gamma}$, i.e. $\left({}^{(2)}\Gamma_{\ep_{1}\ep_{2}}^{\ep_{1}\ep_{2}}\right)^2$. From \Eq{appen:specific2RDM},  
\ba
{}^{(2)}\Gamma_{\ep_{1}\ep_{2}}^{\ep_{1}\ep_{2}}&=\frac{1}{2}\sum_{n} a_{nn}f_n(\ep_1)f_n(\ep_2)(\de_{\ep_1\ep_4}\de_{\ep_2\ep_3}-\de_{\ep_1\ep_3}\de_{\ep_2\ep_4}).
\ea
Thus, 
\ba\label{append:res}
&2\sum_{\ep_{1},\ep_{2}}\left({}^{(2)}\Gamma_{\ep_{1}\ep_{2}}^{\ep_{1}\ep_{2}}\right)^2\nn\\
&=\sum_{\ep_{1},\ep_{2}}\sum_{n,p}\frac{a_{nn}a_{pp}}{2}f_n(\ep_1)f_n(\ep_2)f_p(\ep_2)f_p(\ep_1)(\de_{\ep_1\ep_1}\de_{\ep_2\ep_2}-\de_{\ep_1\ep_2}\de_{\ep_2\ep_1})^2\nn\\
&=\sum_{n,p}\frac{a_{nn}a_{pp}}{2}\left[\left(\sum_{\ep}f_n(\ep)f_p(\ep)\right)^2-\sum_{\ep}f_n(\ep)f_p(\ep)\right].
\ea
Note that this term is identical to the term in the first square bracket of  \Eq{appen:2purity}. Inserting \Eqs{append:res}{appen:2purity} into \Eq{eq:dp2}, we arrive at the final form of the  two-body distilled purity in \Eq{dis2}:
\be
\label{appen:dis2}
\begin{split}
\tilde{P}_{2}&={\sum_{\substack{n,m\\n\neq m}}}^\prime {\sum_{\substack{p,q\\p\neq q}}}^\prime a_{nm}a_{pq}A_2\left[\de_{s_{nm},1}\de_{s_{pq,1}}\de_{\alp_2\de_1}\de_{\bet_1\ga_2}(N-s_{mq})\right.\\
&\left.+A_3(\de_{\bet_1\ga_1}\de_{\bet_2\ga_2}-\de_{\bet_1\ga_2}\de_{\bet_2\ga_1})(\de_{\alp_1\de_1}\de_{\alp_2\de_2}-\de_{\alp_1\de_2}\de_{\alp_2\de_1})\right],
\end{split}
\ee           
where $A_2=f_m(\bet_1)f_q(\de_1)(1-f_m(\alp_1))(1-f_m(\alp_2))(1-f_q(\ga_1))(1-f_q(\ga_2))$ and $A_3=f_m(\bet_2)f_q(\de_2)$.
Both $S$-coherences of order 1 and 2 are captured by $\tilde{P}_2$.

%%%%%%%%%%%%%%%%%%%%%%%%%%%%%%%%%%%%%%%%%%%%%%%%%%%%%%%%%%%%%%%%%%%%%%
\subsubsection{Example}

As an example, consider the 3-particle system with $\h{\rho}_e=\sum_{n,m=1}^4a_{nm}|\Phi_n\ra\la\Phi_m|$ where $|\Phi_1\ra=c^{\da}_1c^{\da}_2c^{\da}_3|0\ra, |\Phi_2\ra=c^{\da}_1c^{\da}_2c^{\da}_6|0\ra, |\Phi_3\ra=c^{\da}_3c^{\da}_4c^{\da}_5|0\ra, |\Phi_4\ra=c^{\da}_4c^{\da}_5c^{\da}_6|0\ra$. In this case, \Eq{appen:2purity} yields
\be
\begin{split}
P_2&=2(a_{11}^2+a_{22}^2+a_{33}^2+a_{44}^2)+(a_{11}+a_{22})^2\\
&+(a_{33}+a_{44})^2+4(|a_{12}|^2+|a_{34}|^2)\\
&+2(|a_{13}|^2+|a_{24}|^2+a_{13}a_{42}+a_{31}a_{24}).
\end{split}
\ee
The associated distilled purity is:
\be
\label{eq:ex2}
\begin{split}
\tilde{P}_2&= 4(|a_{12}|^2+|a_{34}|^2)\\
&+2(|a_{13}|^2+|a_{24}|^2+a_{13}a_{42}+a_{31}a_{24}).
\end{split}
\ee
The reduced purity has contributions from the populations of the Slater determinant states and the $S$-coherences between them, while the distilled purities captures just the $S$-coherences.  The $S$-coherences between the states that differ by one-body transition and two-body transitions contribute to $P_2$ and $\tilde{P}_2$ . For example, the term $|a_{12}|^2$ appears due to a one-body between $|\Phi_1\ra$ and $|\Phi_2\ra$ while $|a_{13}|^2$ appears due to a two-body transition between $|\Phi_1\ra$ and $|\Phi_3\ra$. Moreover, two distinct pairs of states that differ by the same two-body transitions also contribute to the two-body reduced and distilled purities. The term $a_{13}a_{42}$ appears as both pairs of states ($|\Phi_1\ra$ and $|\Phi_3\ra$, and $|\Phi_2\ra$ and $|\Phi_4\ra$) differ by the same two-body transition as $|\Phi_3\ra=c^{\da}_5c_2c^{\da}_4c_1|\Phi_1\ra$ and $|\Phi_4\ra=c^{\da}_5c_2c^{\da}_4c_1|\Phi_2\ra$.

%%%%%%%%%%%%%%%%%%%%%%%%%%%%%%%%%%%%%%%%%%%%%%%%%%%%%%%%%%%%%%%%%%%%%%
\bibliography{Paper}

\end{document}